\documentclass[italian,english]{article}
\usepackage[T1]{fontenc}
\usepackage[latin9]{inputenc}
\usepackage{amstext}
\usepackage{amssymb}
\usepackage{esint}
\usepackage{babel}
\begin{document}

\title{\textbf{On the Jeans theorem and the ``Tolman-Oppenheimer-Volkov
equation'' in $\textbf{\ensuremath{R^{2}}}$ gravity}}

\author{\textbf{$^{1}$Rishabh Jain, $^{2,3}$Burra G. Sidharth and $^{4*}$Christian
Corda}}

\maketitle
\begin{center}
$^{1}$Homer. L. Dodge Department of Physics and Astronomy University
of Oklahoma, Norman , Oklahoma USA , 73019.
\par\end{center}

\begin{center}
\textbf{$^{2}$}International Institute for Applicable Mathematics
\& Information Sciences (IIAMIS),  Hyderabad (India) \& Udine (Italy).\textbf{ }
\par\end{center}

\begin{center}
\textbf{$^{3}$}G. P. Birla Observatory \& Astronomical Research Centre
B. M. Birla Science Centre, Adarsh Nagar, Hyderabad - 500 063 (India). 
\par\end{center}

\begin{center}
$^{4}$Research Institute for Astronomy and Astrophysics of Maragha
(RIAAM), P.O. Box 55134-441, Maragha, Iran.
\par\end{center}

\begin{center}
$^{*}$Correspondence: cordac.galilei@gmail.com
\par\end{center}
\begin{abstract}
Corda, Mosquera Cuesta and Lorduy Gòmez have shown that spherically
symmetric stationary states can be used as a model for galaxies in
the framework of the linearized $\textbf{\ensuremath{R^{2}}}$ gravity.
Those states could represent a partial solution to the Dark Matter
Problem. Here we discuss an improvement of this work. In fact, as
the star density is a functional of the invariants of the associated
Vlasov equation, we show that any of these invariants is in its turn
a functional of the local energy and the angular momentum. As a consequence,
the star density depends only on these two integrals of the Vlasov
system. This result is known as the \textquotedblleft{}\emph{Jeans
theorem}\textquotedblright{}. In addition, we find an analogous of
the historical Tolman-Oppenheimer-Volkov equation for the system considered
in this paper.

For the sake of completeness, in the final Section of the paper we
consider two additional models which argue that Dark Matter could
not be an essential element.
\end{abstract}

\section{Introduction}

The Dark Matter problem started in the 30's of last century \cite{key-1}.
When one observes the Doppler shift of stars which move near the plane
of our Galaxy and calculates the velocities, one finds a large amount
of matter inside the Galaxy which prevents the stars to escape out.
That (supposed and unknown) matter generates a gravitational force
very large, that the luminous mass in the Galaxy cannot explain. In
order to achieve the very large discrepancy, the sum of all the luminous
components of the Galaxy should be two or three times more massive.
On the other hand, one can calculate the tangential velocity of stars
in orbits around the Galactic center like a function of distance from
the center. The result is that stars which are far away from the Galactic
center move with the same velocity independent on their distance out
from the center. 

These strange issues generate a portion of the Dark Matter problem.
In fact, either luminous matter is not able to correctly describe
the radial profile of our Galaxy or the Newtonian Theory of gravitation
cannot describe dynamics far from the Galactic center. 

Other issues of the problem arise from the dynamical description of
various self-gravitating astrophysical systems. Examples are stellar
clusters, external galaxies, clusters and groups of galaxies. In those
cases, the problem is similar, as there is more matter arising from
dynamical analyses with respect to the total luminous matter. 

Zwicky \cite{key-2} found that in the Coma cluster the luminous mass
is too little to generate the gravitational force which is needed
to hold the cluster together \cite{key-2}. 

The more diffuse way to attempt to solve the problem is to assume
that Newtonian gravity holds at all scales that should exist and that
they should exist non-luminous components which contribute to the
missing mass. There are a lot of names which are used to define such
non-luminous components. The MAssive Compact Halo Objects (MACHOs)
are supposed to be bodies composed of normal baryon matter, which
do not emit (or emit little) radiation and drift through interstellar
space unassociated with planetary systems \cite{key-3}. They could
be black holes and/or neutron stars populating the outer reaches of
galaxies. The Weakly Interacting Massive Particles (WIMPs) are hypothetical
particles which do not interact with standard matter (baryons, protons
and neutrons) \cite{key-4}. Hence, they should be particles outside
the Standard Model of Particles Physics but they have not yet been
directly detected. Dark Matter is usually divided in three different
flavors, Hot Dark Matter (HDM) \cite{key-5}, Warm Dark Matter (WDM)
\cite{key-6} and Cold Dark Matter (CDM) \cite{key-7}. HDM should
be composed by ultrarelativistic particles like neutrinos. CDM should
consist in MACHOs, WIMPs and axions, which are very light particles
with a particular behavior of self-interaction \cite{key-8}. If we
consider the standard model of cosmology, the most recent results
from the Planck mission \cite{key-37} show that the total mass\textendash{}energy
of the known universe contains $4.9\%$ ordinary matter, 26.8\% Dark
Matter and 68.3\% Dark Energy.

An alternative approach is to explain large scale structure without
dark components in the framework of Extended Theories of Gravity {[}10-17{]}
and references within. In other words, we call ``Dark Matter'' a
gravitational effect that we do not yet understand as modification
to both Newtonian and Einstenian gravity could be needed, see from
{[}10-17{]}. The underlying idea in Extended Theories of Gravity is
that General Relativity is a particular case of a more general effective
theory which comes from basic principles {[}10-17{]}. The standard
Einstein-Hilbert action of General Relativity \cite{key-17,key-18}
is modified by adding new degrees of freedom, like high order curvature
corrections (the so-called $f(R)$ theories of gravity {[}10-17,20{]}
and scalar fields (the generalization of the Nordstr\"{o}m-Jordan-Fierz-Brans-Dicke
theory of gravitation {[}21-25{]}, which is known as scalar-tensor
gravity {[}26-28{]}). In this different context, one assumes that
gravity is not scale-invariant and takes into account only the ``observed''
ingredients, i.e. curvature and baryon matter. Thus, it is not required
to search candidates for Dark Matter which have not yet been found
{[}10-17,20{]}. In this perspective, the gravitational wave astronomy
should be the ultimate test for the physical consistency of General
Relativity or of any other theory of gravitation \cite{key-9}.

\section{$R^{2}$ theory of gravity and Vlasov system}

In the framework of $f(R)$ theories of gravity, the \textbf{$R^{2}$}
theory is the simplest among the class of viable models with $R^{m}$
terms. Those models support the acceleration of the universe in terms
of cosmological constant or quintessence as well as an early time
inflation \cite{key-10,key-14,key-19}. Moreover, they should pass
the Solar System tests, as they have an acceptable Newtonian limit,
no instabilities and no Brans-Dicke problem (decoupling of scalar)
in scalar-tensor version. We recall that the \textbf{$R^{2}$} theory
was historically proposed in \cite{key-28} with the aim of obtaining
the cosmological inflation. 

The \textbf{$R^{2}$} theory arises from the action \cite{key-29}
\begin{equation}
S=\int d^{4}x\sqrt{-g}(R+bR^{2}+\mathcal{L}_{m}),\label{eq: high order 1}
\end{equation}
where $b$ represents the coupling constant of the $R^{2}$ term.
In general, when the constant coupling of the \textbf{$R^{2}$ }term
in the gravitational action (\ref{eq: high order 1}) is much minor
than the linear term $R\mbox{, }$the variation from standard General
Relativity is very weak and the theory can pass the Solar System tests
\cite{key-29}. In fact, as the effective scalar field arising from
curvature is highly energetic, the constant coupling of the the \textbf{$R^{2}$}
non-linear term $\rightarrow0$ \cite{key-29}. In that case, the
Ricci scalar, which represents an extra dynamical quantity in the
metric formalism, should have a range longer than the size of the
Solar System. This is correct when the effective length of the scalar
field $l$ is much shorter than the value of $0.2\; mm$ \cite{key-30}.
Hence, this effective scalar field results hidden from Solar System
and terrestrial experiments. By analysing the deflection of light
by the Sun in the \textbf{$R^{2}$} theory through a calculation of
the Feynman amplitudes for photon scattering, one sees that, to linearized
order, the result is the same as in standard General Relativity \cite{key-21}.
By assuming that the dynamics of the matter (the stars making of the
galaxy) can be described by the Vlasov system, a model of stationary,
spherically symmetric galaxy can be obtained. This issue was described
in detail in \cite{key-29}, in the framework of the \textbf{$R^{2}$}
theory. For the sake of completeness, in this Section we shortly review
this issue.

In this paper we consider Greek indices run from 0 to 3. By varying
the action of Eq. (\ref{eq: high order 1}) with respect to $g_{\mu\nu}$,
one gets the field equations \cite{key-29}

\begin{equation}
\begin{array}{c}
G_{\mu\nu}+b\{2R[R_{\mu\nu}-\frac{1}{4}g_{\mu\nu}R]+\\
\\
-2R_{;\mu;\nu}+2g_{\mu\nu}\square R\}=T_{\mu\nu}^{(m)}.
\end{array}\;,\label{eq: einstein-general}
\end{equation}
By taking the trace of this equation, the associated Klein - Gordon
equation for the Ricci curvature scalar 

\begin{equation}
\square R=E^{2}(R+T),\label{eq: KG}
\end{equation}
is obtained \cite{key-29}, where $\square$ is the d'Alembertian
operator and the energy term, $E$, has been introduced for dimensional
motivations \cite{key-29}
\begin{equation}
E^{2}\equiv\frac{1}{6b}\;.\label{eq: Eya}
\end{equation}
Hence, $b$ is positive \cite{key-29}. $T_{\mu\nu}^{(m)}$ in Eq.
(\ref{eq: einstein-general}) is the standard stress-energy tensor
of the matter and General Relativity is easily re- obtained for $b=0$
in Eq. (\ref{eq: einstein-general}).

As we study interactions between stars at galactic scales, we consider
the linearized theory in vacuum ($T_{\mu\nu}^{(m)}=0$), which gives
a better approximation than Newtonian theory \cite{key-29}. Calling
$\widetilde{R}$ the linearized quantity which corresponds to $R$,
considering the plane wave \cite{key-29} 
\begin{equation}
b\widetilde{R}=a(\overrightarrow{p})\exp(iq^{\beta}x_{\beta})+c.c.\label{eq: sol S}
\end{equation}
with  \cite{key-29} 
\begin{equation}
\begin{array}{ccc}
p^{\beta}\equiv(\omega,\overrightarrow{p}), &  & \omega=p\equiv|\overrightarrow{p}|\\
\\
q^{\beta}\equiv(\omega_{E},\overrightarrow{p}), &  & \omega_{E}=\sqrt{E^{2}+p^{2}}.
\end{array}\label{eq: k e q}
\end{equation}
one can choose a gauge for a gravitational wave propagating in the$+z$
direction in which a first order solution of eqs. (\ref{eq: einstein-general})
with $T_{\mu\nu}^{(m)}=0$ is given by the the conformally flat line
element \cite{key-29} 
\begin{equation}
ds^{2}=[1+b\widetilde{R}(t,z)](dx^{2}+dy^{2}+dz^{2}-dt^{2}),\label{eq: metrica puramente scalare}
\end{equation}
We recall that the dispersion law for the modes of $b\widetilde{R}$,
i.e. the second of Eq. (\ref{eq: k e q}), is that of a wave-packet
\cite{key-29}. The group-velocity of a wave-packet of $b\widetilde{R}$
centred in $\overrightarrow{p}$ is \cite{key-29}

\begin{equation}
\overrightarrow{v_{G}}=\frac{\overrightarrow{p}}{\omega_{E}}.\label{eq: velocita' di gruppo}
\end{equation}
From the second of Eq. (\ref{eq: k e q}) and Eq. (\ref{eq: velocita' di gruppo})
one gets \cite{key-29}

\begin{equation}
v_{G}=\frac{\sqrt{\omega_{E}^{2}-E^{2}}}{\omega_{E}},\label{eq: velocita' di gruppo 2}
\end{equation}
which can be rewritten as \cite{key-29}

\begin{equation}
E=\sqrt{(1-v_{G}^{2})}\omega_{E}.\label{eq: relazione massa-frequenza}
\end{equation}
If one assumes that the dynamics of the stars making of the galaxy
is described by the Vlasov system, the gravitational forces between
the stars will be mediated by the metric (\ref{eq: metrica puramente scalare}).
Thus, the key assumption is that, in a cosmological framework, the
wave-packet of $b\widetilde{R}$ centred in $\overrightarrow{p}$,
which is given by the (linearized) spacetime curvature, governs the
motion of the stars \cite{key-29}. In this way the ``\emph{curvature}''
energy $E$ is identified as the Dark Matter content of a galaxy of
typical mass-energy $E\simeq10^{45}\: g$, in ordinary c.g.s. units
\cite{key-29}. As $E\simeq10^{45}\: g,$ from Eq. (\ref{eq: Eya})
one gets $b\simeq10^{-34}\: cm^{4}$ in natural units \cite{key-29}.
Hence, the constant coupling of the $\textbf{\ensuremath{R^{2}}}$
term in the action (\ref{eq: high order 1}) is much minor than the
linear term $R\mbox{\mbox{ }}$ and the variation from standard General
Relativity is very weak. This implies that the theory can pass the
Solar System tests as the effective length of the scalar field is
$l\ll0.2\: mm$ \cite{key-29}.

We can use a conformal transformation \cite{key-29,key-33} to rescale
the line-element (\ref{eq: metrica puramente scalare}) like 
\begin{equation}
\tilde{g}_{\alpha\beta}=e^{\Phi}g_{\alpha\beta},\label{eq: conforme}
\end{equation}
where we set \cite{key-29} 
\begin{equation}
\Phi\equiv b\widetilde{R}.\label{eq: trucco conforme 2}
\end{equation}
Thus, in the linearized theory we get

\begin{equation}
e^{\Phi}=1+b\widetilde{R}.\label{eq: trucco conforme}
\end{equation}

Hence, it is the Ricci scalar, i.e. the \emph{scalaron} \cite{key-28,key-29},
the scalar field which translates the analysis into the conformal
frame, the Einstein frame \cite{key-14,key-29}.

Particles in a spacetime make up an ensemble with no collisions and
are governed by a line element like Eq. (\ref{eq: metrica puramente scalare})
if the particle density satisfies the Vlasov equation {[}30,32-34{]}

\begin{equation}
\partial_{t}f+\frac{p^{a}}{p^{0}}\partial_{x^{a}}f-\Gamma_{\mu\nu}^{a}\frac{p^{\mu}p^{\nu}}{p^{0}}\partial_{p^{a}}f=0,\label{eq: Vlasov}
\end{equation}
where $\Gamma_{\mu\nu}^{\alpha}$ are the Christoffel coefficients,
$f\:$ is the particle density, $p^{0}$ is given by $p^{a}$($a=1,2,3$)
according to the relation {[}30,32-34{]}

\begin{equation}
g_{\mu\nu}p^{\mu}p^{\nu}=-1\label{eq: mass-shell}
\end{equation}
 Eq. (\ref{eq: mass-shell}) means that the four momentum $p^{\mu}$
lies on the mass-shell of the spacetime {[}30,32-34{]}. 

In general, the Vlasov-Poisson system is introduced through the system
of equations {[}30,32-34{]}

\begin{equation}
\begin{array}{c}
\partial_{t}f+v\cdot\bigtriangledown_{x}f-\bigtriangledown_{x}U\cdot\bigtriangledown_{v}f=0\\
\\
\bigtriangledown\cdot U=4\pi\rho\\
\\
\rho(t,x)=\int dvf(t,x,v),
\end{array}\label{eq: VP}
\end{equation}
In Eqs. (\ref{eq: VP} )$t$ is the time and $x$ and $v$ are the
position and the velocity of the stars respectively. $U=U(t,x)$ is
the average Newtonian potential generated by the stars. The system
(\ref{eq: VP} ) represents the non-relativistic kinetic model for
an ensemble of particles (stars in the galaxy) with no collisions.
The stars interact only through the gravitational forces which they
generate collectively, are considered as pointlike particles, and
we neglect the relativistic effects {[}30,32-34{]}. The function $f(t,x,v)$
in the Vlasov-Poisson system (\ref{eq: VP}) is non-negative and gives
the density on phase space of the stars within the galaxy \cite{key-29}.

In this approach, the first order solutions of the Klein-Gordon equation
(\ref{eq: KG}) for the Ricci curvature scalar are considered like
galactic high energy \emph{scalarons}, which are expressed in terms
of wave-packets having stationary solutions within the Vlasov system
\cite{key-29}. The energy of the wave-packet is interpreted like
the Dark Matter component which guarantees the galaxy's equilibrium
\cite{key-29}. This approximation is not as precise as one would
aspect \cite{key-29}, but here we consider it as the starting point
of our analysis. 

The analysis in \cite{key-29} permits to rewrite the Vlasov-Poisson
system in spherical coordinates as 

\begin{equation}
-\frac{d^{2}b\widetilde{R}}{dt^{2}}+\frac{1}{r^{2}}\frac{d}{dr}\left(\frac{d}{dr}b\widetilde{R}r^{2}\right)=(1+2b\widetilde{R})\mu(t,r),\label{eq: KG 6}
\end{equation}

\begin{equation}
\mu(t,r)=\int\frac{dp}{\sqrt{1+p^{2}}}f(t,x,p),\label{eq: mass-shell 4}
\end{equation}

\begin{equation}
\partial_{t}f+\frac{p}{\sqrt{1+p^{2}}}\cdot\partial_{x}f-\left[\left(\frac{d}{dt}b\widetilde{R}+\frac{x\cdot p}{\sqrt{1+p^{2}}}\frac{1}{r}\frac{d}{dr}b\widetilde{R}\right)p+\frac{x}{\sqrt{1+p^{2}}}\frac{1}{r}\frac{d}{dr}b\widetilde{R}\right]\cdot\partial_{p}f=0,\label{eq: Vlasov 4}
\end{equation}
In Eqs. (\ref{eq: KG 6}), (\ref{eq: mass-shell 4}) and (\ref{eq: Vlasov 4})
$p$ denotes the vector $p=(p_{1},p_{2},p_{3})$ with $p^{2}=|p|^{2},$
and $x$ denotes the vector $x_{i}=(x_{1},x_{2},x_{3})$ \cite{key-29}
.

As one is interested in stationary states, one calls $\lambda$ the
wavelength of the \emph{``galactic'' gravitational wave} (\ref{eq: metrica puramente scalare}),
i.e. the characteristic length of the gravitational perturbation \cite{key-29}.
One further assumes that $\lambda\gg d,$ $d$ being the galactic
scale of order $d\thicksim10^{5}$ light-years \cite{key-35,key-36}.
In other words, the gravitational perturbation can be considered ``\emph{frozen-in}''
with respect to the galactic scale \cite{key-29}. 

In that way, one can write down the system of equations defining the
stationary solutions of eqs. (\ref{eq: KG 6}), (\ref{eq: mass-shell 4})
and (\ref{eq: Vlasov 4}) \cite{key-29}

\begin{equation}
\frac{1}{r^{2}}\frac{d}{dr}\left(\frac{d}{dr}b\widetilde{R}r^{2}\right)=(1+2b\widetilde{R})\mu(r),\label{eq: KG 7}
\end{equation}

\begin{equation}
\mu(r)=\int\frac{dp}{\sqrt{1+p^{2}}}f(x,p),\label{eq: mass-shell 5}
\end{equation}

\begin{equation}
p\cdot\partial_{x}f-\frac{1}{r}\frac{d}{dr}b\widetilde{R}[(p\cdot x)p+x]\cdot\partial_{p}f=0,\label{eq: Vlasov 5}
\end{equation}
Therefore, the idea in \cite{key-29} is that the spin-zero degree
of freedom arising from the $R^{2}$ term in the gravitational Lagrangian,
i.e. the scalaron, is a potential candidate for the dark matter. In
this approach, the dominant contribution to the curvature within a
galaxy comes from the scalaron field equation (\ref{eq: KG}). That
equation has a proper baryon source term. This enables the baryons
themselves to evolve obeying a collisionless Boltzmann equation. Then
the baryons can propagate on the curvature generated by the scalaron.

\section{The Jeans theorem and the \textquotedblleft{}Tolman-Oppenheimer-Volkov
equation''}

The following results will be obtained adapting the ideas introduced
in {[}32-34,37{]}. Let us start by recalling some important definitions
in the conformal frame:

\begin{equation}
\begin{array}{ccc}
P(r)\equiv\int\frac{dp}{\sqrt{1+p^{2}}}\left(\frac{x\cdot p}{r}\right)^{2}f(x,p) &  & radial\: pressure\\
\\
\rho(r)\equiv(1+2bR)\int dp\sqrt{1+p^{2}}f(x,p) &  & mass-energy\: density\\
\\
P_{T}(r)\equiv\int\frac{dp}{\sqrt{1+p^{2}}}\left|\frac{x\wedge p}{r}\right|f(x,p) &  & tangential\: pressure.
\end{array}\label{eq: definizioni}
\end{equation}
Thus, Eq. (\ref{eq: KG 7}) becomes 
\begin{equation}
\frac{1}{r^{2}}\frac{d}{dr}\left(\frac{d}{dr}b\widetilde{R}r^{2}\right)=\rho(r)-P(r)-2P_{T}(r).\label{eq: diventa}
\end{equation}

Let us consider the stationary solution of our stellar dynamics model,
i.e. Eqs. (\ref{eq: KG 7}), (\ref{eq: mass-shell 5}) and (\ref{eq: Vlasov 5}).
The particle density is a functional of the invariants of the Vlasov
equation (14). The Jeans theorem states that any of these invariants
must be a functional of the local energy and the angular momentum.
In that way, the particle density depends only on these two integrals
of the system under consideration. In order to prove these statements,
one introduces the new coordinates 
\begin{equation}
\begin{array}{c}
r=\left|x\right|\\
\\
Y=\left(1+\frac{b}{2}\widetilde{R}\right)\frac{x\cdot p}{\left|x\right|}\\
\\
Z=\left(1+b\widetilde{R}\right)\left[\left|x\right|^{2}-\left|p\right|^{2}-\left(x\cdot p\right)^{2}\right].
\end{array}\label{eq: new coordinates}
\end{equation}
We note that, $f$ being spherically symmetric, one can write it as
a function of $r,Y,Z$, i.e. $f(x,p)\rightarrow f(r,Y,Z)$. Thus,
Eq. (\ref{eq: Vlasov 5}) in the new coordinates reads 
\begin{equation}
Y\partial_{r}f+\left[\frac{Z}{r^{3}}-\left(1+b\widetilde{R}\right)\frac{d}{dr}b\widetilde{R}\right]\partial_{Y}f=0.\label{eq: interessante}
\end{equation}
This equation has the same form of eq. (2.11) in \cite{key-42} with
$m(r)=\left(1+b\widetilde{R}\right)\left(\frac{d}{dr}b\widetilde{R}r^{2}\right).$
Thus, by applying the result in \cite{key-42}, one obtains that $f$
must have the form 
\begin{equation}
f(r,Y,Z)=A\left(\bar{E},Z\right),\label{eq: forma f}
\end{equation}
where 
\begin{equation}
\bar{E}(r,Y,Z)=\frac{1}{2}Y^{2}+\frac{1}{2}\frac{Z}{r^{2}}+\frac{1+b\widetilde{R}}{2}=\left(\frac{1+b\widetilde{R}}{2}\right)\left(1+p^{2}\right).\label{eq: E barra}
\end{equation}
Returning to the system of Eqs. (\ref{eq: KG 7}), (\ref{eq: mass-shell 5})
and (\ref{eq: Vlasov 5}), one gets immediately 
\begin{equation}
f(x,p)=A(E,X),\label{eq:  f di x p}
\end{equation}
where 
\begin{equation}
E=\left(1+\frac{b}{2}\widetilde{R}\right)\sqrt{1+p^{2}}\label{eq: energia locale}
\end{equation}
is the local energy of the particles and 
\begin{equation}
X=\left(1+\frac{b}{2}R\right)\left|x\wedge p\right|^{2}\label{eq: momento angolare locale}
\end{equation}
is the modulus squared of the local angular momentum. Using Eq. (\ref{eq:  f di x p})
and a transformation of variables, the system (\ref{eq: definizioni})
becomes 
\begin{equation}
\begin{array}{c}
P(r)=\frac{\pi}{r^{2}}\int dE\int dX\, A(E,X)\sqrt{E^{2}-\frac{X}{r^{2}}-\left(1+\frac{b}{2}\widetilde{R}\right)}\\
\\
\rho(r)=\frac{\pi}{r^{2}}\int dE\, E^{2}dX\,\frac{A(E,X)}{\sqrt{E^{2}-\frac{X}{r^{2}}-\left(1+\frac{b}{2}\widetilde{R}\right)}}\\
\\
P_{T}(r)=\frac{\pi}{2r^{4}}\int dE\int dX\,\frac{X\, A(E,X)}{\sqrt{E^{2}-\frac{X}{r^{2}}-\left(1+\frac{b}{2}\widetilde{R}\right)}}.
\end{array}\label{eq: ridefinite}
\end{equation}
A direct computation permits to obtain 
\begin{equation}
\frac{d}{dr}P(r)=-\left(1+b\widetilde{R}\right)\frac{d}{dr}\rho(r)-\frac{2}{r}\left[P(r)-P_{T}(r\right],\label{eq: Tolman-Oppenheimer-Volkov}
\end{equation}
which is the analogous of the historical Tolman-Oppenheimer-Volkov
equation \cite{key-43,key-44} for the system under analysis in this
paper. 

For the sake of completeness, we stress that the Jeans instability
(gravitational stability) has been analysed in the context of modified
theories of gravity for rotating/non-rotating configurations in {[}40-42{]}.
Related questions on the Tolman-Oppenheimer-Volkov equation and the
Jeans instability (but not for plane waves) were also studied in {[}43-45{]}.
The Tolman-Oppenheimer-Volkov equation has been also analysed in the
dimensional gravity's rainbow in the presence of cosmological constant
in \cite{key-63} and in the framework of dilaton gravity in \cite{key-64}.

\section{Two additional models}

For the sake of completeness, we play the devil's advocate and consider
two models which argue that Dark Matter is not an essential element,
even though popular models postulate that it comprises roughly a fourth
of a universe. Our starting point is the relation \cite{key-45,key-46}
\begin{equation}
G=G_{0}\left(1-\frac{t}{t_{0}}\right)\label{e15}
\end{equation}
where $G_{0}$ is the present value of $G$, $t_{0}$ is the present
age of the universe and $t$ the time elapsed from the present epoch.
Similarly one could deduce that \cite{key-46} 
\begin{equation}
r=r_{0}\left(\frac{t_{0}}{t_{0}+t}\right).\label{e16}
\end{equation}
In this scheme the gravitational constant $G$ varies slowly with
time. This is suggested by Sidharth's 1997 cosmology \cite{key-45},
which correctly predicted a Dark Energy driven accelerating universe
at a time when the accepted paradigm was the Standard Big Bang cosmology
in which the universe would decelerate under the influence of dark
matter. 

We reiterate the following: The problem of galactic rotational curves
\cite{key-46,key-47}. We would expect, on the basis of straightforward
dynamics that the rotational velocities at the edges of galaxies would
fall off according to 
\begin{equation}
v^{2}\approx\frac{GM}{r}.\label{e33}
\end{equation}
However it is found that the velocities tend to a constant value,
\begin{equation}
v\sim300\: km/sec.\label{e34}
\end{equation}
This as known had lead to the postulation of as yet undetected additional
matter, the so called Dark Matter. We observe that from Eq. (\ref{e16})
it can be easily deduced that \cite{key-48}
\begin{equation}
a\equiv(\ddot{r}_{0}-\ddot{r})\approx\frac{1}{t_{0}}(t\ddot{r_{0}}+2\dot{r}_{0})\approx-2\frac{r_{0}}{t_{0}^{2}},\label{e35}
\end{equation}
as we are considering infinitesimal intervals $t$ and nearly circular
orbits. Equation ($\ref{e35}$) shows that there is an anomalous inward
acceleration, as if there is an extra attractive force, or an additional
central mass \cite{key-49}.

Thus,
\begin{equation}
\frac{GMm}{r^{2}}+\frac{2mr}{t_{0}^{2}}\approx\frac{mv^{2}}{r}\label{e36}
\end{equation}
From Eq. (\ref{e36}) it follows that 
\begin{equation}
v\approx\left(\frac{2r^{2}}{t_{0}^{2}}+\frac{GM}{r}\right)^{1/2}\label{e37}
\end{equation}
Eq. ($\ref{e37}$) shows that at distances within the edge of a typical
galaxy, that is $r<10^{23}\: cm,$ the equation ($\ref{e33}$) holds
but as we reach the edge and beyond, that is for $r\geq10^{24}\: cm,$
we have $v\sim10^{7}\:\frac{cm}{sec}$, in agreement with eq. ($\ref{e34}$).

Then, the time variation of $G$ explains observation without invoking
dark matter. It may also be mentioned that other effects like the
Pioneer anomaly and shortening of the period of binary pulsars can
be deduced \cite{key-50}, while new effects also are predicted.

Milgrom \cite{key-51} approached the problem by modifying Newtonian
dynamics at large distances. This approach is purely phenomenological.
The idea was that perhaps standard Newtonian dynamics works at the
scale of the solar system but at galactic scales involving much larger
distances, the situation might be different. However a simple modification
of the distance dependence in the gravitation law, as pointed by Milgrom
would not do, even if it produced the asymptotically flat rotation
curves of galaxies. Such a law would predict the wrong form of the
mass velocity relation. So Milgrom suggested the following modification
to Newtonian dynamics: A test particle at a distance $r$ from a large
mass $M$ is subject to the acceleration $a$ given by 
\begin{equation}
a^{2}/a_{0}=MGr^{-2},\label{em1}
\end{equation}
where $a_{0}$ is an acceleration such that standard Newtonian dynamics
is a good approximation only for accelerations much larger than $a_{0}$.
The above equation however would be true when $a$ is much less than
$a_{0}$. Both statements can be combined in the heuristic relation
\begin{equation}
\mu(a/a_{0})a=MGr^{-2}.\label{em2}
\end{equation}
In Eq. ($\ref{em2}$); $\mu(x)\approx1$ when $x\gg1,$ and, $\mu(x)\approx x$
when $x\ll1$. It is worthwhile to note that ($\ref{em1}$) or ($\ref{em2}$)
are not deduced from any theory, but rather are an ad hoc fit to explain
observations. Interestingly it must be mentioned that most of the
implications of Modified Newton Dynamics (MOND) do not depend strongly
on the exact form of $\mu$.

It can then be shown that the problem of galactic velocities is solved
{[}55-59{]}.

It is interesting to note that there is an interesting relationship
between the varying $G$ approach, which has a theoretical base and
the purely phenomenological MOND approach. Let us write 
\begin{equation}
\beta\frac{GM}{r}=\frac{r^{2}}{t_{0}^{2}}\,\mbox{or}\,\beta=\frac{r^{3}}{GMt_{0}^{2}}.\label{eq: Final}
\end{equation}
Hence, 
\begin{equation}
\alpha_{0}=v^{2}/r=\frac{GM}{r^{2}}\,\alpha=\frac{r}{t_{0}^{2}}.\label{eq: Whence}
\end{equation}
So that 
\begin{equation}
\frac{\alpha}{\alpha_{0}}=\frac{r^{3}}{GMt_{0}^{2}}=\beta\label{eq: Finalissima}
\end{equation}
At this stage we can see a similarity with MOND. For if $\beta\ll1$
we are with the usual Newtonian dynamics and if $\beta>1$ then we
get back to the varying $G$ case exactly as with MOND.

\section{Concluding remarks}

The results in \cite{key-29} have shown that spherically symmetric
stationary states can be used as a model for galaxies in the framework
of the linearized $\textbf{\ensuremath{R^{2}}}$ gravity. Those states
could, in principle, be a partial solution to the Dark Matter Problem.
In this paper, an improvement of this work has been discussed. As
the star density is a functional of the invariants of the associated
Vlasov equation, it has been shown that any of these invariants is
in turn a functional of the local energy and the angular momentum.
Then, the star density depends only on these two integrals of the
Vlasov system. This result represents the so called \textquotedblleft{}Jeans
theorem\textquotedblright{}. In addition, an analogous of the historical
Tolman-Oppenheimer-Volkov equation \cite{key-43,key-44} for the system
considered in this paper has been discussed. We tried this extension
of previous work in \cite{key-29} because, on one hand, the Jeans
theorem is important in galaxy dynamics and in the framework of molecular
clouds \cite{key-65}. On the other hand, the historical Tolman-Oppenheimer-Volkov
equation constrains the structure of a spherically symmetric body
of isotropic material which is in static gravitational equilibrium,
as modelled by metric theories of gravity, starting from the general
theory of relativity \cite{key-43,key-44}. Thus, a viable extended
theory of gravity, like the $\textbf{\ensuremath{R^{2}}}$ gravity,
must show consistence with these two important issues.

For the sake of completeness, in Section 4 of this paper, two additional
models which argue that Dark Matter could not be an essential element
have been discussed. In fact, Dark Matter is considered a mysterious
and controversial issue. There is indeed a minority of researchers
who think that the dynamics of galaxies could not be determined by
massive, invisible dark matter halos, see \cite{key-9,key-10} and
{[}51-59{]}. We think that, at the present time, there is not a final
answer to the Dark Matter issue. In other words, it is undoubtedly
true that the Universe exhibits a plethora of mysterious phenomena
for which many unanswered questions still exist. Dark Matter is an
important part of this intriguing puzzle. Thus, when one works in
classical, modern, and developing astrophysical and cosmological theories,
it is imperative to repeatedly question their capabilities, identify
possible shortcomings, and propose corrections and alternative theories
for experimental submission. In the procedures and practice of scientific
professionals, no such clues, evidence, or data may be overlooked. 

Finally, we take the chance to stress that important future impacts
which could help to a better understanding of the important Dark Matter
issue could arise from the nascent gravitational wave astronomy \cite{key-66}.
The first direct detection of gravitational waves by the LIGO Collaboration,
the so called event GW150914 \cite{key-66}, represented a cornerstone
for science and for astrophysics in particular. We hope that the gravitational
wave astronomy will become an important branch of observational astronomy
which will aim to use gravitational waves to collect observational
data not only about astrophysical objects such as neutron stars, black
holes and so on, but also about the mysterious issues of Dark Matter
and Dark Energy. In order to achieve this prestigious goal, a network
including interferometers with different orientations is required
and we're hoping that future advancements in ground-based projects
and space-based projects will have a sufficiently high sensitivity
{[}61-63{]}.

For the benefits of the reader, we also signal two important works
on self-gravitating systems \cite{key-69} and on Jeans mass for anisotropic
matter \cite{key-70}.

\section{Conflict of Interests }

The authors declare that there is no conflict of interests regarding
the publication of this paper.

\section{Acknowledgements}

The authors thank an unknown referee for useful comments. Christian
Corda has been supported financially by the Research Institute for
Astronomy and Astrophysics of Maragha (RIAAM), Research Project No.
1/4717-113.

\end{document}